\documentclass[prl,twocolumn,nofootinbib, preprintnumbers, superscriptaddress]{revtex4-1}
\usepackage{amsmath,amssymb,amscd,simplewick}
\usepackage{listings}
\usepackage{dsfont}
\usepackage{slashed}
\usepackage{color}
\usepackage{ulem}
\usepackage{float}

\usepackage{graphicx}
\usepackage{epstopdf}
\usepackage{subfigure}
\usepackage{epsfig}

\usepackage{xcolor}
\usepackage[colorlinks=true,
            linkcolor=blue,
            urlcolor=blue,
            citecolor=green,          
            bookmarks=true,
            bookmarksnumbered=true,
            breaklinks=true,
            pdfstartview=FitBH]{hyperref}

\hypersetup{pdfauthor = {Shao-Feng Ge},
	     pdftitle = {}, 
	     pdfsubject = {}, 
             pdfkeywords = {}, 
	     pdfcreator = {LaTeX with hyperref package},
	     pdfproducer = {dvips + ps2pdf} }

\definecolor{gesfpurple}{rgb}{0.47,0.19,0.42}

\definecolor{gesflanse}{rgb}{0.00,0.50,0.50}

\definecolor{gesfblue}{rgb}{0.08,0.42,0.76}

\definecolor{gesfred}{rgb}{1,0,0}

\definecolor{gesfwhite}{rgb}{1,1,1}

\definecolor{gesfblack}{rgb}{0,0,0}

\definecolor{gesfxd}{rgb}{0,0,1}

\newcommand{\geqn} [1]{\hypersetup{linkcolor=blue}(\ref{#1})\hypersetup{linkcolor=blue}}
\newcommand{\gfig} [1]{{\hypersetup{linkcolor=violet}Fig.~\ref{#1}\hypersetup{linkcolor=blue}}}

\definecolor{Orange}{cmyk}{0,0.61,0.87,0}
\definecolor{JungleGreen}{cmyk}{0.99,0,0.52,0}
\definecolor{OliveGreen}{cmyk}{0.64,0,0.95,0.40}
\definecolor{Brown}{cmyk}{0,0.81,1,0.60}
\definecolor{RoyalBlue}{cmyk}{0.71,0.53,0,0.12}
\definecolor{Gray}{cmyk}{0,0,0,0.40}
\definecolor{LightPink}{cmyk}{0.0,0.25,0,0}
\definecolor{LLightPink}{cmyk}{0.0,0.10,0,0}
\definecolor{LightBlue}{cmyk}{0.25,0,0,0}
\definecolor{LightGray}{cmyk}{0,0,0,0.2}

\begin{document}

\title{Probing the Dark Axion Portal with Muon Anomalous Magnetic Moment}
\author{Shao-Feng Ge}
\email{gesf@sjtu.edu.cn}
\affiliation{Tsung-Dao Lee Institute \& School of Physics and Astronomy, Shanghai Jiao Tong University, Shanghai 200240, China}
\affiliation{Key Laboratory for Particle Astrophysics and Cosmology (MOE) \& Shanghai Key Laboratory for Particle Physics and Cosmology, Shanghai Jiao Tong University, Shanghai 200240, China}

\author{Xiao-Dong Ma}
\email{maxid@sjtu.edu.cn}
\affiliation{Tsung-Dao Lee Institute \& School of Physics and Astronomy, Shanghai Jiao Tong University, Shanghai 200240, China}
\affiliation{Key Laboratory for Particle Astrophysics and Cosmology (MOE) \& Shanghai Key Laboratory for Particle Physics and Cosmology, Shanghai Jiao Tong University, Shanghai 200240, China}

\author{Pedro Pasquini}
\email{ppasquini@sjtu.edu.cn}
\affiliation{Tsung-Dao Lee Institute \& School of Physics and Astronomy, Shanghai Jiao Tong University, Shanghai 200240, China}
\affiliation{Key Laboratory for Particle Astrophysics and Cosmology (MOE) \& Shanghai Key Laboratory for Particle Physics and Cosmology, Shanghai Jiao Tong University, Shanghai 200240, China}

\begin{abstract}
We propose a new scenario of using the dark axion portal at one-loop level
to explain the recently observed muon anomalous magnetic moment
by the Fermilab Muon g-2 experiment. Both axion/axion-like particle (ALP)
and dark photon are involved in the same vertex with photon.
Although ALP or dark photon alone cannot explain muon $g-2$,
since the former provides only negative contribution while
the latter has very much constrained parameter space, dark
axion portal can save the situation and significantly extend
the allowed parameter space. The observed
muon anomalous magnetic moment 
provides a robust probe of the dark axion portal scenario.
\end{abstract}

\maketitle 

{\bf Introduction} --
The muon anomalous magnetic moment $a_\mu \equiv (g_\mu-2)/2$,
where $g_\mu$ is the muon $g$-factor, is one of the most
precisely measured physical parameters in the Standard Model
(SM) of particle physics
\cite{Miller:2012opa, Jegerlehner:2017gek,Zyla:2020zbs}.
The Muon g-2 experiment \cite{Muong-2:2015xgu, Keshavarzi:2019bjn} at Fermilab
provides the currently best measurement \cite{Fermilab21}
\begin{eqnarray}
  a_\mu^{\rm exp} ({\rm FNAL})
=
  116592040\,(54)\times 10^{-11},
\label{eq:Fermilab}
\end{eqnarray}
which is consistent with the previous measurement
$116 592 080\,(63)\times 10^{-11}$
\cite{Bennett:2006fi}
by the E821 experiment
at Brookhaven National Laboratory (BNL).
Then the world average becomes
\begin{eqnarray}
  a_\mu^{\rm exp}
=
  116592061\,(41) \times 10^{-11}.
\label{eq:FermilabBNL}
\end{eqnarray}
From the BNL result to the Fermilab one,
both the central value and the uncertainty
decreases.

Huge amount of work has been done to match the
unprecedented precision.
The SM contribution to $a_\mu$ 
contains four parts \cite{Keshavarzi:2019abf},
\begin{eqnarray}
    a^{\rm SM}_\mu 
  =
    a_\mu^{\rm QED}
  +
    a_\mu^{\rm EW} 
  +
    a_\mu^{\rm HVP}
  +
    a_\mu^{\rm HLbL}
  \,.
\end{eqnarray}
The first two are the QED and electroweak (EW) 
predictions, respectively, while
$a_\mu^{\rm HVP}$ is the hadronic vacuum polarization 
(HVP) and $a_\mu^{\rm HLbL}$ the hadronic light-by-light 
(HLbL) contribution.
Although the biggest source of uncertainty comes from the hadronic
part \cite{Hagiwara:2003da, Gerardin:2020gpp,Chao:2021tvp,Borsanyi:2020mff},
the most recent calculations \cite{Blum:2018mom,Blum:2019ugy,Davier:2019can}
have included the updated measurement of the hadronic
contributions
\cite{Xiao:2017dqv, Lees:2018dnv,Hoferichter:2019mqg}.
The latest theoretical calculation
\cite{Aoyama:2020ynm} gives
\begin{eqnarray}
  a_\mu^{\rm SM}
=
  116591810\,(43) \times 10^{-11},
\label{eq:theory_result}
\end{eqnarray}
where the uncertainty mainly comes from 
the hadronic vacuum polarization $a_\mu^{\rm HVP}$
and the light-by-light part $a_\mu^{\rm HLbL}$.

The longstanding discrepancy \cite{Benayoun:2015gxa, Jegerlehner:2017gek}
between theoretical predictions
\cite{Jegerlehner:2009ry, Aoyama:2020ynm} and experimental
results is also observed by the new measurement
\geqn{eq:FermilabBNL} at Fermilab with $4.2\,\sigma$
significance (combined with BNL E821),
\begin{eqnarray}
  \Delta a_\mu
\equiv 
  a_\mu^{\rm exp} 
-
  a_\mu^{\rm SM}
=
  251 \, (59)\times 10^{-11} \,.
\label{eq:discrepancy}
\end{eqnarray}
The discrepancy increases from $3.7\,\sigma$ to
$4.2\,\sigma$ from the BNL measurement to the new
world average.

Due to its unprecedented precision, the muon
anomalous magnetic moment provides a sensitive probe of
new physics (NP) beyond the SM \cite{Czarnecki:2001pv,Yin:2020afe,Capdevilla:2021rwo}.
The previous $3.7\,\sigma$ discrepancy between the E821
measurement and the SM prediction has stimulated many novel ideas. An incomplete list 
includes lepton flavor violation \cite{Lindner:2016bgg},
$Z'$ \cite{Gninenko:2001hx,Baek:2001kca,Ma:2001md,Altmannshofer:2016brv},
neutral scalars \cite{Crivellin:2010ty,Chen:2015vqy,Abu-Ajamieh:2018ciu,Jana:2020pxx},
ALP \cite{Marciano:2016yhf}, leptoquarks \cite{Chakraverty:2001yg,Cheung:2001ip}, 
supersymmetry \cite{Grifols:1982vx,Barbieri:1982aj,Martin:2001st,Stockinger:2006zn,Padley:2015uma,Belyaev:2016oxy,Endo:2019bcj,Kpatcha:2019pve},
dark photon \cite{Fayet:2007ua,Pospelov:2008zw,TuckerSmith:2010ra,Mohlabeng:2019vrz,Fabbrichesi:2020wbt},
and dark matter portals \cite{Agrawal:2014ufa,Belanger:2015nma,Kowalska:2017iqv,Calibbi:2018rzv,Kawamura:2020qxo,Jana:2020joi}.

In this letter, we explore the possibility that the dark axion portal \cite{Kaneta:2016wvf}
with coupling among ALP $a$, photon $\gamma$,
and a massive dark photon $\gamma'$
can explain the observed muon anomalous magnetic moment at the Fermilab
Muon g-2 experiment. Since this dimension-5 operator was proposed only recently
and involves two invisible particles, its coupling $C_{a \gamma \gamma'}$
is not strongly constrained yet. With TeV scale new physics,
$C_{a \gamma \gamma'} \sim 3 / \mbox{TeV}$, a sizable parameter
space is still available as we will elaborate in this letter.

\begin{figure} [t]
\includegraphics[scale = 0.75]{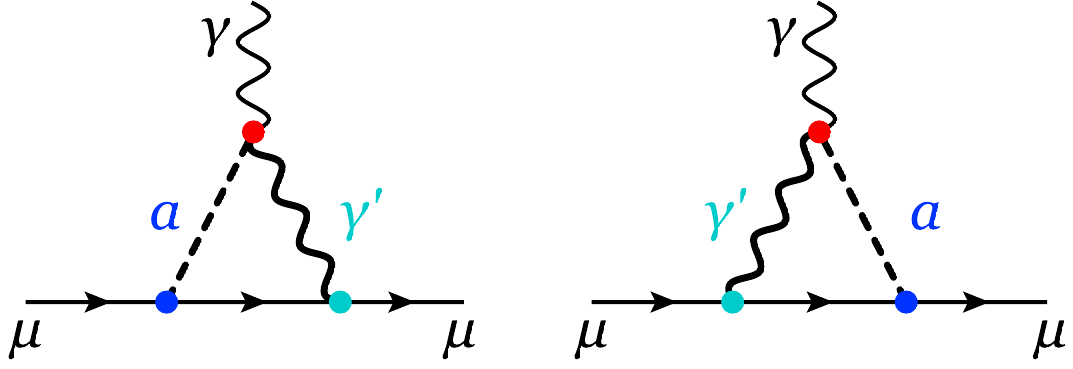}
\caption{
The one-loop contribution to the muon anomalous magnetic moment
from the dark axion portal that couples photon ($\gamma$),
ALP ($a$), and dark photon ($\gamma'$).
}
\label{fig:darkAxionPortal}
\end{figure}

{\bf The Dark Axion Portal Contribution} --
The dark axion portal \cite{Kaneta:2016wvf} establishes the
connection between the visible sector with the dark one via not just
a single ALP or a single photon but both of them,
\begin{eqnarray}
  \mathcal L
\ni
  \frac 1 2 C_{a \gamma \gamma'} a F^{\mu \nu} \widetilde X_{\mu \nu} \,,
\label{eq:darkAxionPortal}
\end{eqnarray}
where $F_{\mu \nu} \equiv \partial_\mu A_\nu - \partial_\nu A_\mu$
is the photon field strength. Through this dimension-5
operator, the CP-violating ALP $a$ couples with the dual
field strength of dark photon
$\widetilde X_{\mu \nu} \equiv \frac 1 2 \epsilon_{\mu \nu \alpha \beta} X^{\alpha \beta}$
where
$X_{\mu \nu} \equiv \partial_\mu X_\nu - \partial_\nu X_\mu$.
As shown in \gfig{fig:darkAxionPortal}, the dark axion portal
can contribute to the muon anomalous magnetic moment if the
ALP and dark photon also couple with muon, 
\begin{eqnarray}
  \mathcal L
\ni
  y^\mu_a a \bar \mu (i \gamma_5) \mu 
- \epsilon e \bar \mu \gamma^\nu \mu X_\nu.
\label{eq:axionDPmuon}
\end{eqnarray}
Here, $y_a^\mu$ is the Yukawa coupling with ALP while
$\epsilon$ is the kinetic mixing between the dark photon
and photon, $\frac 1 2 \epsilon F_{\mu\nu} X^{\mu\nu}$.
In principle, the ALP coupling with two photons can also
contribute by replacing the dark photon in \gfig{fig:darkAxionPortal}
with a photon \cite{Barr-Zee,Marciano:2016yhf}.
However, due to its stringent constraint,
we omit this diagram for simplicity.

The contribution of dark axion portal depicted in \gfig{fig:darkAxionPortal}
is divergent. With the cut-off regularization, the result can be
expressed in terms of the ultra-violet (UV) scale $\Lambda$,
\begin{subequations}
\begin{eqnarray}
  a_\mu
=
  \frac{m_\mu }{4\pi^2}
  \epsilon y^\mu_a C_{a \gamma \gamma'} G,
\label{eq:darkAxionPortala}
\end{eqnarray}
where the loop function $G$ is
\begin{eqnarray}
  G
\equiv
  \int_0^1 dx
& \hspace{-2mm}
\Biggl[
  (1-x)
\left(
  \ln{\frac{\Lambda^2}{(1-x)m_a^2+x^2 m_\mu^2}}
- \frac 1 2
\right)
\nonumber
\\
- &
  \frac{(1-x) m^2_{\gamma'} + 2 x^2 m_\mu^2}{ m_a^2-m_{\gamma'}^2}
  \ln \frac{(1-x)m_a^2+x^2 m_\mu^2}{ (1-x) m_{\gamma'}^2+x^2 m_\mu^2}
\Biggr],
\end{eqnarray}
\label{eq:darkAxionPortalGa}
\end{subequations}
as a function of the ALP mass $m_a$, the dark photon mass $m_{\gamma'}$,
and the muon mass $m_\mu$. 
The dark axion portal contribution \geqn{eq:darkAxionPortala} has linear dependence on the Yukawa coupling $y_a^\mu$. When $y_a^\mu$ is larger than the SM
counterpart $m_\mu / v$ where $v \approx 246\,\mbox{GeV}$ is the Higgs vacuum expectation value, the dark axion portal contribution can be enhanced in comparison with
the SM one. Similar feature has been observed and
named as
chiral enhancement in many models but with quadratic dependence \cite{Kannike:2011ng, Dermisek:2013gta,Crivellin:2018qmi}.

Since the interaction in \geqn{eq:darkAxionPortal}
is non-renormalizable, it is only valid up to 
some cut-off scale $\sim C^{-1}_{a \gamma \gamma'}$. In addition, the divergent
loop integral can be assumingly regularized by
a similar cut-off $\Lambda$ with origin from the same UV physics. 
However, the predicted $a_\mu$ in \geqn{eq:darkAxionPortala}
has mild dependence on $\Lambda$ which only appears in
a log function. Orders of variation in $\Lambda$ can only
change $a_\mu$ by several times which can be easily
compensated by tuning couplings. For comparison, the Yukawa
coupling $y^\mu_a$ and the kinetic mixing parameter $\epsilon$
are dimensionless, hence cannot directly reflect the new physics
scale. 

The dark axion portal can also contribute
to the muon anomalous magnetic moment at two-loop
level via the photon vacuum polarization sub-diagram
\cite{deNiverville:2018hrc}. However, this
contribution is always negative and numerically 
negligible. For example, with
$C_{a\gamma\gamma'} = 3\,\mbox{TeV}^{-1}$,
the two-loop contribution is roughly two orders of magnitude smaller
than its one-loop counterpart. Therefore, we neglect
their contribution and focus on the  one-loop diagrams
in this letter. Note that the dark axion portal
coupling,
$C_{a\gamma\gamma'} = 3\,\mbox{TeV}^{-1}$,
adopted here satisfies existing experimental
constraints \cite{deNiverville:2018hrc}.

In principle, the ALP can also couple with the SM
$Z$ boson,
$\frac 1 2 C_{a\gamma Z} a F_{\mu \nu} \tilde Z^{\mu \nu}$
\cite{Alonso-Alvarez:2018irt,Bauer:2018uxu}.
Then a similar contribution from the ALP-photon-$Z$
vertex, by replacing the dark photon $\gamma'$ in
\gfig{fig:darkAxionPortal} with $Z$.
The analytical formula \geqn{eq:darkAxionPortalGa}
still applies after replacing the dark photon mass
$m_{\gamma'}$ by the $Z$ boson mass $m_Z$, the
coupling constants $C_{a\gamma\gamma'}$ by
$C_{a\gamma Z}$ and $e \epsilon$ by
$g_V=\frac{g}{c_W}\left(\frac 1 4-s_W^2\right)\approx
4.5\times 10^{-2}e$ of the vector part of the
$Z$-muon coupling while the axial-vector part does
not contribute due to the mismatch of parity and charge conjugation
properties. The good thing is that the $Z$ coupling with
muon and the $Z$ boson mass have already been
measured, hence reducing the number of parameters
by two. However, the current bound on the coupling
$C_{a \gamma Z}\lesssim 0.03\,{\rm TeV}^{-1}$
\cite{Cheung:2008ii} from the anomalous $Z$ decay
$Z\to \gamma a$ \cite{Jaeckel:2015jla} is
rather stringent. The contribution from the
ALP-photon-$Z$ vertex is negligibly small.

\begin{figure} [t]
\centering
\includegraphics[scale = 0.75]{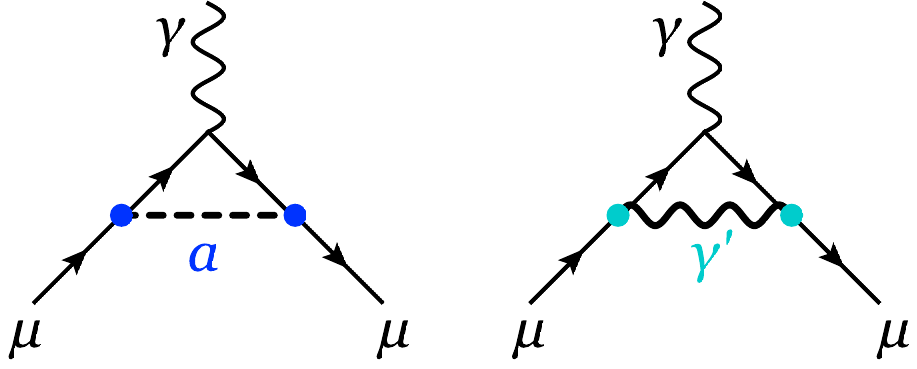}
\caption{The ALP (left) and dark photon (right)
contributions to the muon anomalous magnetic moment.}
\label{fig:axionDP}
\end{figure}

{\bf The Individual Contribution of ALP or Dark Photon} --
The ALP or dark photon alone can also contribute to the
muon anomalous magnetic moment as shown in \gfig{fig:axionDP}.
We first consider the contribution from the ALP
which is finite,
\begin{subequations}
\begin{eqnarray}
  a_{\mu}^{a}
& = &
  \frac{(y_{a}^\mu)^2}{4\pi^2}
  \frac{m_\mu^2}{ m_a^2}
  F_a\left(\frac{m_\mu}{ m_a}\right),
\\
  F_a(\eta)
& \equiv &
  -\frac 1 2
  \int_{0}^{1} \mathrm{~d} x 
  \frac{x^3 }{ (1-x)\left(1-\eta^2 x\right)+\eta^2 x}.
\end{eqnarray}
\label{eq:ALP}
\end{subequations}
Note that this ALP-only contribution is negative
\cite{Marciano:2016yhf} since $F_a(\eta)\lesssim 0$
where $\eta \equiv m_\mu / m_a$.
The pseudoscalar case is completely different from
the scalar scenario which can contribute a positive
term. A pseudoscalar alone cannot explain why
the observed $a^{\rm exp}_\mu$ is larger than the
SM prediction $a^{\rm SM}_\mu$ unless the experimental
measurement is smaller than the theoretical
prediction.

The contribution from the dark photon shown in the
right panel of \gfig{fig:axionDP} takes the form as
\cite{Fayet:2007ua, Pospelov:2008zw,TuckerSmith:2010ra},
\begin{subequations}
\begin{eqnarray}
  a_\mu^{\gamma'}
& = &
  \frac{\epsilon^2 e^2}{4\pi^2}
  \frac{m_\mu^2}{m_{\gamma'}^2}
  F_{\gamma'}\left(\frac{m_\mu}{ m_{\gamma'}}\right),
\\
  F_{\gamma'}(\eta)
& \equiv &
  \frac 1 2
  \int_{0}^{1} \mathrm{~d} x 
  \frac{2x^2(1-x) }{ (1-x)\left(1-\eta^2 x\right)+\eta^2 x},
\end{eqnarray}
\label{eq:darkPhoton}
\end{subequations}
with $\eta \equiv m_\mu / m_{\gamma'}$.
Different from the loop function $F_a$ for the ALP-only
contribution, $F_{\gamma'} \gtrsim 0$ always holds.
It seems that the single contribution from the dark photon
can explain the observed muon anomalous magnetic moment.
However, the required parameter space in the $m_{\gamma'} - \epsilon$
plane to explain the observed $\Delta a_\mu$ has
already been excluded by other experimental bounds
\cite{Fabbrichesi:2020wbt} as we will discuss in
detail below.

\begin{figure} [t]
\centering
\includegraphics[scale = 0.22]{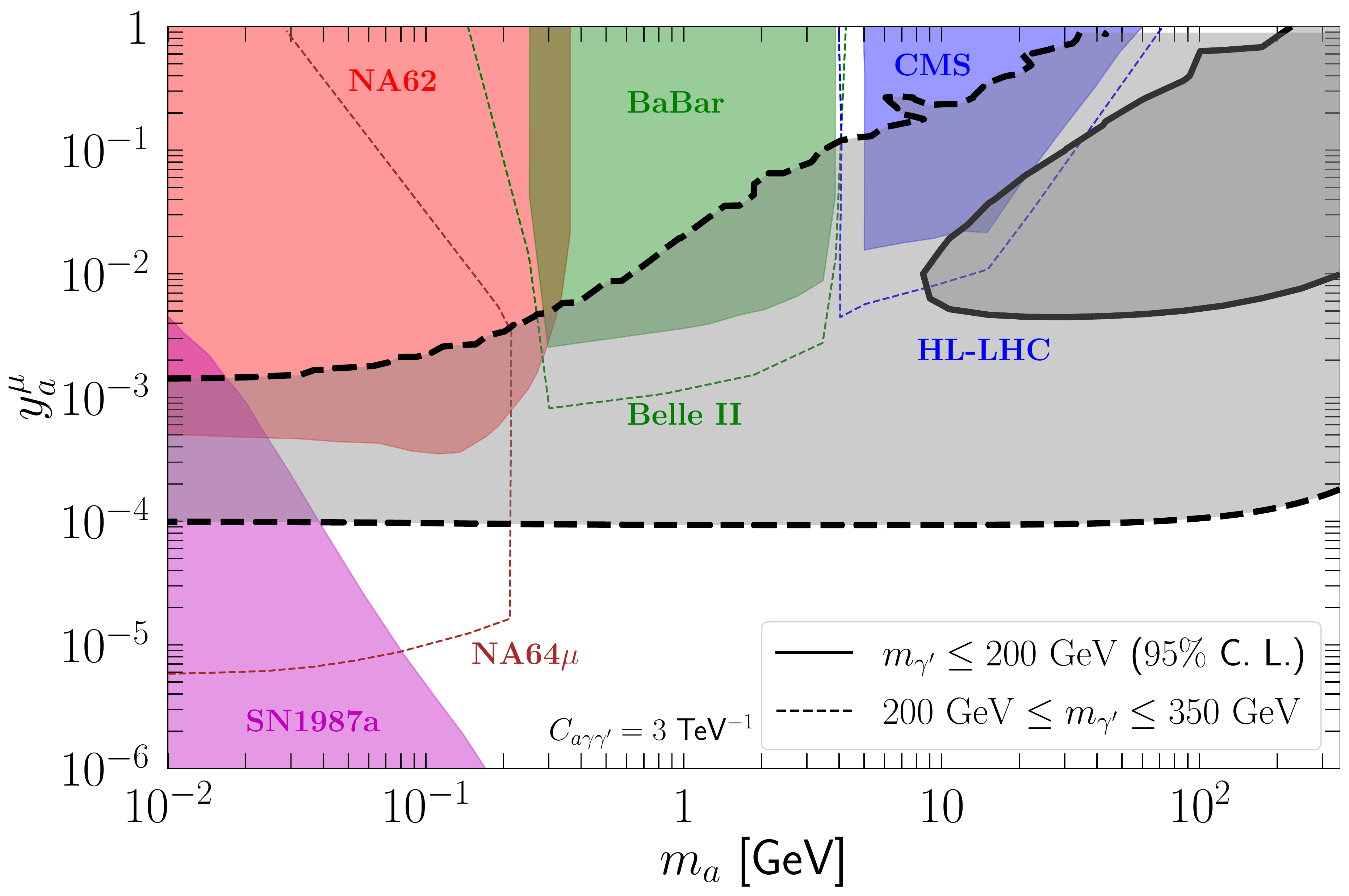}
\caption{
Experimental constraints on the ALP Yukawa coupling with muon from
1) supernova cooling from SN1987a (purple region) \cite{Croon:2020lrf};
2) kaon decay $K \rightarrow \mu \nu a$ \cite{Batell:2016ove}
(red region) at NA62;
3) final-state radiation $e^+ e^- \rightarrow \mu^+ \mu^- a$ (green region)
at the BaBar experiment \cite{TheBABAR:2016rlg, Batell:2016ove};
4) rare $Z$ decay (blue region) at CMS \cite{Sirunyan:2018nnz}.
The projected sensitivities at NA64 (brown dashed line) \cite{Chen:2017awl},
Belle II (green dashed line), and HL-LHC (blue dashed line) \cite{Batell:2017kty}
are also shown for comparison.
The black contours are the allowed region for explaining the
observed muon anomalous magnetic moment together with dark axion portal
   at 95\% C.L.
}
\label{fig:ALP_bounds}
\end{figure}

{\bf Parameter Space} --
It is instructive to compare the three distinct contributions:
\geqn{eq:darkAxionPortala} for dark axion portal,
\geqn{eq:ALP} for ALP, and \geqn{eq:darkPhoton} for dark photon, 
\begin{subequations}
\begin{eqnarray}
  \frac{a_\mu }{ a_\mu^a}
& \sim &
  \frac{\epsilon }{ y_a^\mu}
  \frac{m_a^2 C_{a \gamma \gamma'} }{  m_\mu}
\sim
  \frac{\epsilon }{ 10^{-3}}
  \frac{0.1}{y_a^\mu}
\left( \frac{m_a}{ 100\,\mbox{GeV}} \right)^2,
\\
  \frac{a_\mu }{ a_\mu^{\gamma'}}
& \sim &
  \frac{y_a^\mu }{ \epsilon e^2}
  \frac{m_{\gamma'}^2 C_{a \gamma \gamma'}}{ m_\mu}
\sim
  10^5\frac{10^{-3}}{ \epsilon}
  \frac{y_a^\mu}{0.1}
  \left(\frac{m_{\gamma'}}{100\,\mbox{GeV}} \right)^2,
\qquad
\end{eqnarray}
\end{subequations}
where $C_{a\gamma\gamma'} \sim {\rm TeV}^{-1}$.
The loop function ratios,
$G / F_a$ and $G / F_{\gamma'}$, are dropped out
since they are comparable with each other.
For the dark axion portal contribution to dominate,
$a_\mu \gg a^a_\mu, a^{\gamma'}_\mu$, the ALP
and dark photon masses are bounded from below,
$m_a \gg \sqrt{y^\mu_a / \epsilon}\, 10\,\mbox{GeV}$ and
$m_{\gamma'} \gg \sqrt{\epsilon / y^\mu_a}\, \sqrt{10}\,\mbox{GeV}$.
If the two dimensionless couplings are comparable with each other,
$\epsilon \sim y^\mu_a$, both the ALP mass $m_a$
and the dark photon mass $m_{\gamma'}$ are around
the GeV scale. Richer mass patterns can be realized
by tuning the two couplings to change the mass limits,
or allowing the ALP-only and dark photon-only
contributions in \gfig{fig:axionDP} to be comparable with
the dark axion portal one in \gfig{fig:darkAxionPortal}.
Below we explore the allowed parameter spaces in detail. 

As argued above, the interesting ALP mass is around GeV 
to a few hundreds of GeV scale.
In this range, the experimental constraints
\cite{Batell:2016ove,Chen:2017awl,Batell:2017kty}
mainly come from SN1987a,
beam dump experiment NA62,
low energy electron positron colliders such as BaBar,
and collider searches at LHC.
We summarize in \gfig{fig:ALP_bounds} those constraints that can apply to
the configuration in this letter, where the ALP only
couples with muon rather than electron or tau. 

The purple region at the left-bottom corner of
\gfig{fig:ALP_bounds} is excluded by the supernova (SN)
cooling rate from the SN1987a observation \cite{Croon:2020lrf}.
The bound reaches $y_a^\mu \approx 10^{-6}$ and can
exclude the mass region up to 0.2\,GeV.

The red region in \gfig{fig:ALP_bounds} is
excluded by the NA62 experiment \cite{Batell:2016ove}
using the search of rare kaon decay channel
$K \rightarrow \mu \nu a$. 
The probe of $m_a$ is limited by the 
kaon mass ($\sim 494\,\mbox{MeV}$), explaining why the
excluded region can only extend to $\lesssim 400$ MeV.  
For comparison, we also show the projected sensitivity
from the $\mu^+N\to \mu^+ N+a$ process
at the NA64 muon beam dump experiment
(brown dashed line) \cite{Chen:2017awl}.
\begin{figure} [t]
  \centering
  \includegraphics[scale = 0.22]{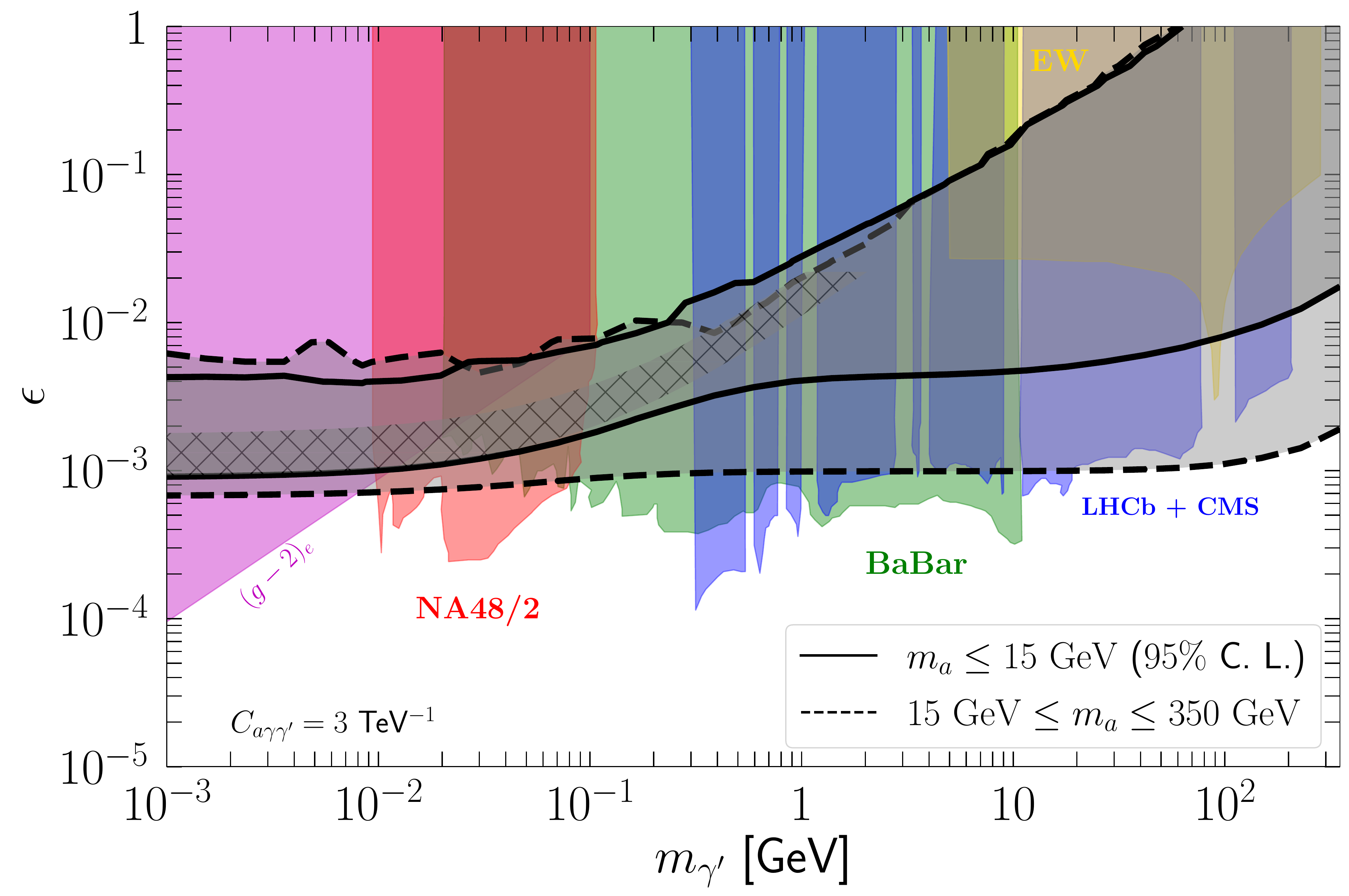}
  \caption{
   Experimental constraints on the dark photon kinetic mixing
   parameter $\epsilon$ as a function of the dark photon mass
   $m_{\gamma '}$ from : 1) electron anomalous magnetic moment
   $(g-2)_e$ (purple) \cite{Pospelov:2008zw}; 2) resonant production
   of dark photon at the BaBar experiment (green) \cite{Lees:2014xha};
   3) dark photon from pion decay at the NA48 experiment (red) \cite{Batley:2015lha}; 
   4) dark photon production from various mesons at LHCb
\cite{Ilten:2016tkc,Aaij:2019bvg}
   and Higgs at CMS (blue) \cite{Sirunyan:2019wqq};
   5) electroweak precision observables (yellow)
    \cite{Curtin:2014cca}.
   The hashed band is the allowed region for explaining the muon g-2
   with dark photon alone while the black contour is obtained
   together with dark axion portal at 95\% C.L.
  }
  \label{fig:DarkPhoton_bounds}
\end{figure}

The green region in \gfig{fig:ALP_bounds}
was obtained from the BaBar experiment
with final-state radiation of the invisible $Z'$ which
further decays into a $\mu^+ \mu^-$ pair,
$e^+e^- \rightarrow  \mu^- \mu^+ Z' \rightarrow \mu^- \mu^+ \mu^- \mu^+$.
Although this bound is not originally obtained for
ALP, it can be easily converted \cite{TheBABAR:2016rlg, Batell:2016ove}.
The BaBar experiment is an electron-positron collider with
center-of-mass energy around $10$ GeV. Consequently, the 
sensitive mass region is $m_a \in [0.1, 4]$ GeV.
The similar situation happens for the CMS rare $Z$ decay
constraint \cite{Sirunyan:2018nnz} shown as the blue region
in \gfig{fig:ALP_bounds} which covers the range from
5\,GeV up to 50\,GeV. For comparison, we also show the
projected sensitivities at Belle II (green dashed line)
and HL-LHC \cite{Batell:2017kty} (blue dashed line).

It is evident that there is a very large region of 
the parameter space to be explored for masses above
0.2\,GeV. All other constraints compiled in
\cite{Batell:2016ove,Chen:2017awl,Batell:2017kty}
involves Yukawa couplings with either electron or
tau leptons and hence cannot apply to the configuration
considered in this letter.
The gap between the BaBar and CMS region can 
be covered by the future HL-LHC searches (blue dashed line)
and Belle II can also increase the sensitivity below 4 GeV
(green dashed lines) \cite{Batell:2017kty}.
Even so, the available parameter space is still quite
sizable.

The current bounds \cite{Fabbrichesi:2020wbt,Filippi:2020kii}
on the dark photon mass $m_{\gamma'}$
and its kinetic mixing $\epsilon$ with photon has been
summarized in \gfig{fig:DarkPhoton_bounds}.
These constraints covers the dark photon mass range 
from 1\,MeV to 1\,TeV. In contrast to the ALP case,
since the kinetic mixing leads to
universal coupling between the dark photon and all charged
leptons, those experimental constraints involving electron
can also apply here. It is interesting to see that the
electron anomalous magnetic moment, $(g-2)_e$, excludes
the very light dark photon scenario (purple region)
\cite{Pospelov:2008zw}. The dark photon can be resonantly
produced at the BaBar experiment.
With 10\,GeV center-of-mass energy, the excluded region
(green) spans from around 20\,MeV up to roughly 10\,GeV \cite{Lees:2014xha}.
In between, the NA48 searches for dark photon from pion decay
cover the red region from 9\,MeV to 100\,MeV \cite{Batley:2015lha}.
The decays of various mesons at LHCb \cite{Ilten:2016tkc,Aaij:2019bvg}
and Higgs at CMS \cite{Sirunyan:2019wqq} also give strong constraints
shown as gaped blue regions. At CMS, the dark photon
is produced from Higgs decay, $h \rightarrow Z \gamma'$
and $h \rightarrow \gamma' \gamma'$, and it further
decays to a pair of muons.
The yellow region comes from the electroweak (EW)
precision observables \cite{Curtin:2014cca}.
It is interesting to observe that the EW precision
observables fill the gap around $Z$ boson mass.

{\bf Revival with Dark Axion Portal} --
With the available parameter space compiled in
\gfig{fig:ALP_bounds} for ALP and
\gfig{fig:DarkPhoton_bounds} for dark photon,
we are ready to explore the allowed region for
explaining the recently observed muon anomalous magnetic
moment at Fermilab. This can be quantitatively done with $\chi^2$
function,
\begin{equation}
  \chi^2
\equiv
\left(
  \frac {\Delta a_\mu - \Delta a^{\rm NP}_\mu}
				{\sigma(\Delta a_\mu)}
\right)^2,
\end{equation}
with central value $\Delta a_\mu$ and uncertainty
$\sigma(\Delta a_\mu)$ taken from
\geqn{eq:discrepancy}.
The new physics prediction $\Delta a^{\rm NP}_\mu$
here contains the four parameters, $m_a$ and $y^\mu_a$
for ALP as well as $m_{\gamma'}$ and $\epsilon$ for
dark photon, in addition to the fixed coupling
$C_{a \gamma \gamma'} = 3\,\mbox{TeV}^{-1}$
and cut-off $\Lambda = 1\,\mbox{TeV}$.
To illustrate the 
 allowed parameter space 
 of ALP, for each point
of \gfig{fig:ALP_bounds} the values of $m_a$ and
$y^\mu_a$ are fixed while
the dark photon parameters
$m_{\gamma'}$ and 
$\epsilon$ are varied to obtain the smallest value
$\chi^2_{\rm min} (m_a, y^\mu_a)$.
The dark 
photon parameters scan
in the range $m_{\gamma'} \in [10^{-3}, 350]$ GeV and $\epsilon \in [10^{-5}, 1]$. However, those points that fall inside 
the experimentally excluded regions of \gfig{fig:DarkPhoton_bounds}
are not included in the scan.
The resulting $\chi^2_{\rm min}(m_a, y^\mu_a)$ is
then a marginalized $\chi^2$ function of just the two ALP parameters.
Similar procedures can also produce a marginalized
$\chi^2_{\rm min}(m_{\gamma'}, \epsilon)$ after scanning
the ALP parameters in the range of $m_{a} \in [10^{-2}, 350]$~GeV
and $y_a^\mu \in [10^{-6}, 1]$ but deducting experimentally 
excluded regions.

So from \gfig{fig:ALP_bounds} we can
read off the values of $m_a$ and $y^\mu_a$, but not the
corresponding values of $m_{\gamma'}$ and $\epsilon$.
The similar situation happens for \gfig{fig:DarkPhoton_bounds}.
The black contours of $m_a$ and $y^\mu_a$ in \gfig{fig:ALP_bounds}
are obtained with $\chi^2_{\rm min}(m_a, y^\mu_a) < 5.99$ at 95\% C.L. and similarly for \gfig{fig:DarkPhoton_bounds}.

The black contours in \gfig{fig:ALP_bounds} cover a large
part of the remaining parameter space. As argued at the
beginning of this section, the ALP mass is bounded from
below,
$m_a \gg \sqrt{y^\mu_a / \epsilon}\, 10\,\mbox{GeV}$,
in order for the dark axion portal contribution to
dominate. With smaller Yukawa coupling $y^\mu_a$, the
ALP mass can also be smaller and hence cover the whole
mass range in \gfig{fig:ALP_bounds}. The ALP
contribution is always
negative \cite{Marciano:2016yhf} no matter what is the
sign of the Yuakwa coupling $y^\mu_a$ as indicated by
\geqn{eq:ALP}. This forbids the possibility of using
only ALP to explain the observed positive $\Delta a_\mu$.
However, in the presence of dark axion portal, the
ALP or pseudoscalar at large receives significant
parameter space to explain the muon anomalous magnetic
moment. Although the black
contour is marginalized over the dark photon parameters,
namely its mass $m_{\gamma'}$ and kinetic mixing $\epsilon$,
we can still show the dependence on the dark photon mass
by specifying the mass range of dark photon to be
marginalized over. The solid black
contour is obtained with $m_{\gamma'} \leq 200\,\mbox{GeV}$
while the dashed one with 
$200\,\mbox{GeV} \leq m_{\gamma'} \leq 350\,\mbox{GeV}$.
It is interesting to see that with larger dark photon mass,
the allowed ALP parameter space becomes larger with the
Yukawa coupling $y^\mu_a$ touching down to as small as
$10^{-4}$. The ALP solution can be readily saved by the
dark axion portal.

For the dark photon parameter space illustrated in
\gfig{fig:DarkPhoton_bounds}, almost all mass range
below 200\,GeV has been experimentally constrained
to $\epsilon \lesssim 10^{-3}$. Especially, the
required parameter space, the hashed region in
\gfig{fig:DarkPhoton_bounds}, for dark photon to
explain the muon anomalous magnetic moment has
been excluded by various observations including
electron $(g-2)_e$, NA48/2, BaBar, and LHCb+CMS.
It is very interesting to see that the dark axion
portal coupling can also help to save the situation.
Now the required parameter space significantly
expands to the black contours, the solid one for
$m_a \leq 15\,\mbox{GeV}$ and the dashed one for
$15\,\mbox{GeV} \leq m_a \leq 350\,\mbox{GeV}$.
The heavy mass region, $m_{\gamma'} \gtrsim
\mathcal O(10)\,\mbox{GeV}$,
still has sizable space. Even the low mass region
around $m_{\gamma'} \approx 10\,\mbox{MeV}$ opens
for $15\,\mbox{GeV} \leq m_a \leq 350\,\mbox{GeV}$.

\begin{figure}[t]
    \centering
    \includegraphics[scale = 0.22]{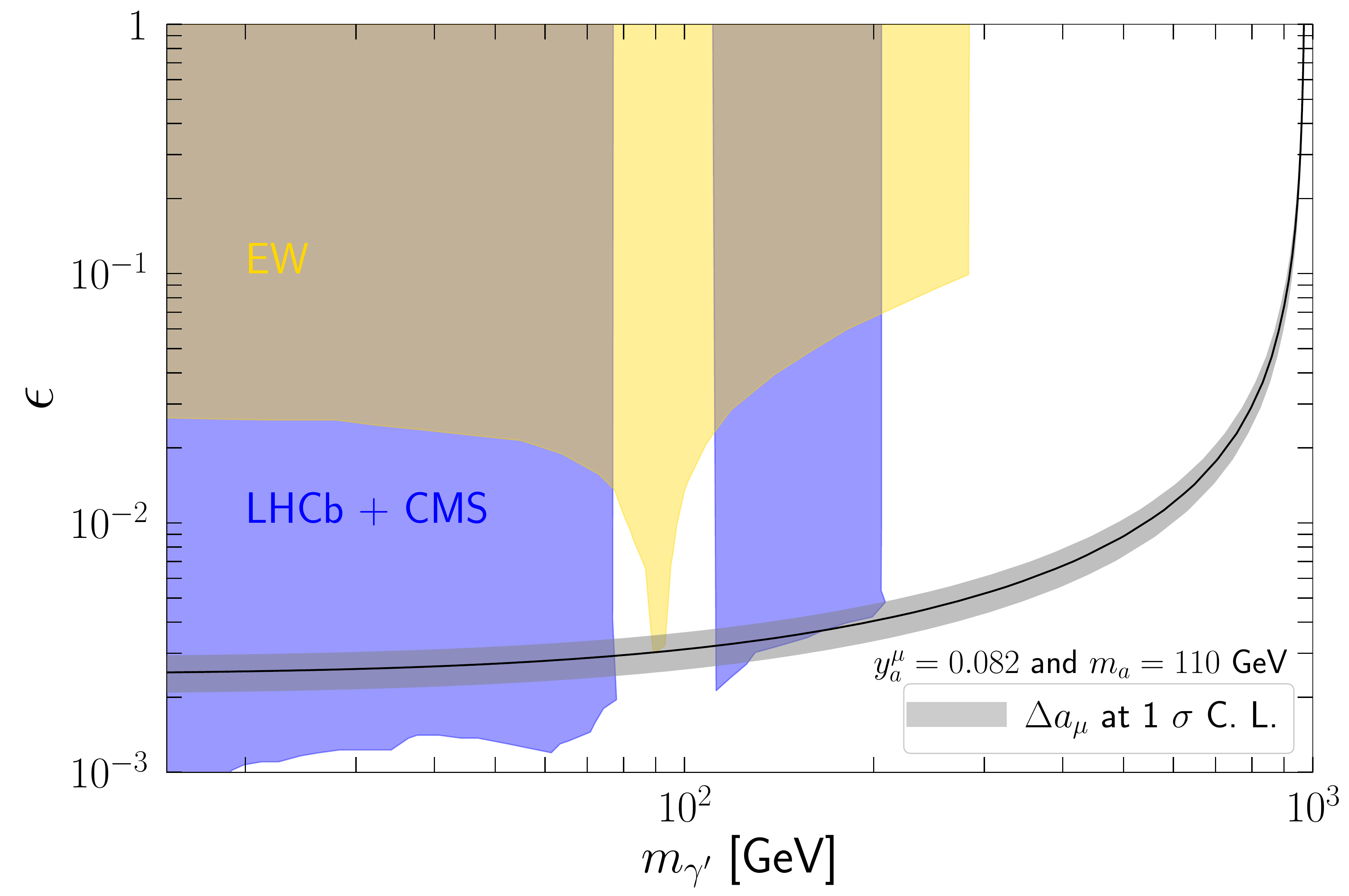}
    \caption{The allowed region for explaining the muon $g-2$ with the fixed ALP parameters $y_a^\mu=0.082$ and $m_a=110$ GeV at 1 $\sigma$ C.L.. }
\label{fig:decoupling_dark_photon}
\end{figure}

In the large mass limit 
($m_a, m_{\gamma '} \gg m_\mu$),
the total contribution shows decoupling features.
To make it explicit, 
we fix the ALP parameters,
$m_a = 110$ GeV and $y_a^\mu = 0.082$,
as an example. The total contribution $\Delta a_\mu(m_{\gamma'},\epsilon)$ is then a function of the two dark
photon parameters $m_{\gamma'}$
and $\epsilon$. \gfig{fig:decoupling_dark_photon} shows
the allowed region of $\Delta a_\mu(m_{\gamma'}, \epsilon)
= (251 \pm 59) \times 10^{-11}$
as grey area. With larger dark photon mass, the dark photon
coupling $\epsilon$ also needs to increase to maintain the
prediction of $\Delta a_\mu(m_{\gamma'}, \epsilon)$.
Otherwise, for a fixed $\epsilon$, the predicted
$\Delta a_\mu(m_{\gamma'}, \epsilon)$ would decrease with the dark photon mass.
This decoupling behavior can be
understood analytically with the approximate forms of \geqn{eq:darkAxionPortalGa}, \geqn{eq:ALP} and \geqn{eq:darkPhoton} 
in the large mass limit,
\begin{subequations}\label{eqn:largemasslimit_all}
\begin{eqnarray}
    a_\mu
& \approx &
 \frac{m_\mu}{4\pi^2} \epsilon y^\mu_a  C_{a\gamma \gamma'}
  \frac{m_a^2 \ln\frac{\Lambda }{ m_a} - m_{\gamma'}^2 \ln \frac{\Lambda }{ m_{\gamma'} }}{m_a^2 - m_{\gamma'}^2},
\label{eqn:largemasslimit_amu}
\\ 
  a_\mu^{a}
& \approx  &
-
  \left(\frac{y^\mu_a }{2\pi}\right)^2
  \frac{m^2_\mu}{m_a^2}
\left(\ln \frac {m_a} {m_\mu} - \frac {11}{12} \right),
\label{eqn:largemasslimit_amua}
\\
  a_\mu^{\gamma'}
& \approx &
  \frac 1 3
 \left(\frac{\epsilon e}{2\pi}\right)^2 
 \frac{m_\mu^2}{m_{\gamma'}^2}.
\label{eqn:largemasslimit_amuga}
\end{eqnarray}
\end{subequations}
For $m_{\gamma'} \gg m_a$, we can see that $a_\mu \propto \ln \frac \Lambda {m_{\gamma'}}$ and
$a^{\gamma'}_\mu \propto 1 / m^2_{\gamma'}$. Both terms
decrease with $m_{\gamma'}$ and hence have decoupling
behavior.
\begin{figure}[t]
    \centering
    \includegraphics[scale = 0.22]{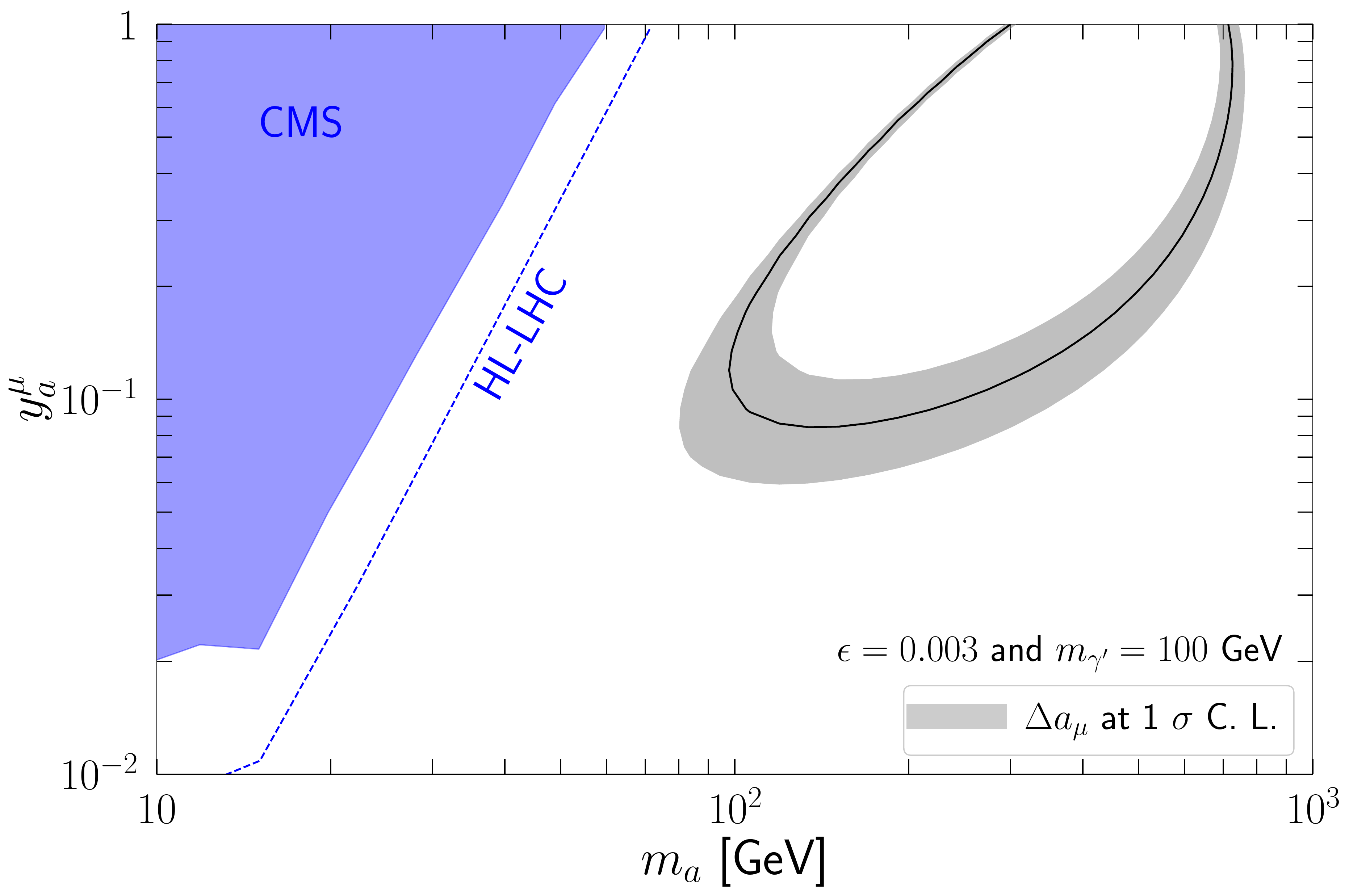}
    \caption{The allowed ALP region for explaining the muon $g-2$ with the fixed dark photon parameters $\epsilon=0.003$ and $m_{\gamma'}=100$ GeV at 1 $\sigma$ C.L..}
\label{fig:decoupling_axion}
\end{figure}

\gfig{fig:decoupling_axion} shows the decoupling behavior in
the ALP parameter space.
Taking the two dark photon parameters, $m_{\gamma'} = 100$ GeV and $\epsilon = 0.003$, 
the allowed region of $\Delta a_{\mu}(m_a, y_a^\mu) = 
(251 \pm 59)\times 10^{-11}$ in the $m_a-y_a^\mu$ plane is
the grey area. 
Since $y_a^\mu \sim \mathcal O(10^{-1})$ is relatively large,
both $a_\mu^a$ and $a_\mu$ provide the dominant contributions,
one has quadratic dependence on $y_a^\mu$ and the other linear.
Consequently, for a given ALP mass $m_a$, there are two possible
solutions for the Yukawa coupling $y^\mu_a$ to match the 
experimental result. This is why the grey area of
\gfig{fig:decoupling_axion} is a circle rather than a line
as in \gfig{fig:decoupling_dark_photon}.

For $m_a \gg m_{\gamma'}$, we can also see decoupling features:
with larger ALP mass the prediction $\Delta a_\mu(m_a, y_a^\mu)$ becomes
smaller. This is because both 
$a_\mu \propto \ln \frac \Lambda {m_a}$
and $a^a_\mu \approx (m^2_\mu / m^2_a) \ln (m_a / m_\mu)$
decrease with $m_a$. With larger ALP mass, the Yukawa coupling
$y^\mu_a$ should also increase in order to maintain the same
prediction of $\Delta a_\mu(m_a, y_a^\mu)$.

However, with both dark photon and ALP parameters, the
decoupling features are not transparent.
For illustration, we take $m_{\gamma'} = 100$ GeV, 
$m_a = 110$ GeV, and $\epsilon = 3\times 10^{-3}$
(combined with our assumptions: $C_{a\gamma\gamma'} = 3$ TeV$^{-1}$ and $\Lambda = 1$ TeV), leading to, 
\begin{eqnarray}
  a_\mu
\approx
  4.2\times 10^{-8} y_a^\mu,
\quad
  a^a_\mu 
\approx
- 1.4 \times 10^{-7} (y_a^\mu)^2,
\end{eqnarray}
together with $a_\mu^{\gamma'} \ll \Delta a_\mu$.
Then either $y_a^\mu = 0.082$ or $y_a^\mu = 0.22$ gives
the observed discrepancy.
The points 
$(m_a, y_a^\mu) = (110~{\rm GeV}, 0.082 \mbox{ or } 0.22)$ 
and $(m_{\gamma'}, \epsilon) = (100~{\rm GeV}, 3\times 10^{-3})$ are
inside our 95\% C. L. band in \gfig{fig:ALP_bounds} and \gfig{fig:DarkPhoton_bounds},
as expected, but $y^\mu_a$ and $\epsilon$
need not to be of the same order.
In other words, the presence of 
both contributions can in fact enlarge the 
parameter space, including larger values of $m_a$ and
$m_{\gamma'}$.

{\bf Conclusion} --
The latest measurement of the muon anomalous
magnetic moment at the Fermilab Muon g-2
experiment further enhances the discrepancy
with theoretical
prediction from 3.7\,$\sigma$ to 4.2\,$\sigma$.
This clearly indicates that there is something new
beyond the SM, although a decisive conclusion
still awaits more data. On one hand, this
discrepancy enhances theoretical exploration
of possible solutions. But on the other, some
solutions have been already excluded, including
either the ALP or dark photon scenario.
Even though the dark axion portal was originally
motivated as a way to connect the visible world
with the dark side, it can surprisingly save
the ALP and dark photon for explaining the
muon anomalous magnetic moment. Since dark matter
contributes five times more energy density in the Universe
than the ordinary matter and the latter already
has rich particle spectrum, there is no reason to
assume that the dark sector is composed of a single
particle. In this sense, the dark axion portal
provides a more interesting option than just ALP or
dark photon. And muon anomalous magnetic moment
can provide a robust probe of this new scenario.

\section*{Acknowledgements}

The authors are supported by the Double First Class start-up fund (WF220442604), the Shanghai Pujiang Program (20PJ1407800), and National Natural Science Foundation of China (No. 12090064). This work is also supported in part by Chinese Academy of Sciences Center for Excellence in Particle Physics (CCEPP).


\begin{thebibliography}{99}

\bibitem{Miller:2012opa}
  J.~P.~Miller, E.~de Rafael, B.~L.~Roberts and D.~St\"ockinger,
  ``{\it Muon (g-2): Experiment and Theory},''
  \href{http://dx.doi.org/10.1146/annurev-nucl-031312-120340}
  {Ann. Rev. Nucl. Part. Sci. \textbf{62}, 237-264 (2012)}

\bibitem{Jegerlehner:2017gek}
F.~Jegerlehner,
``{\it The Anomalous Magnetic Moment of the Muon},''
\href{http://dx.doi.org/10.1007/978-3-319-63577-4}
{Springer Tracts Mod. Phys. \textbf{274} (2017), pp.1-693}.

\bibitem{Zyla:2020zbs}
P.A.~Zyla \textit{et al.} [Particle Data Group],
``{\it Review of Particle Physics},''
\href{http://dx.doi.org/10.1093/ptep/ptaa104}
{PTEP \textbf{2020}, no.8, 083C01 (2020)}

\bibitem{Muong-2:2015xgu}
J.~Grange \textit{et al.} [Muon g-2],
``{\it Muon (g-2) Technical Design Report},''
[\href{http://arxiv.org/abs/1501.06858}{arXiv:1501.06858} [physics.ins-det]].

\bibitem{Keshavarzi:2019bjn}
A.~Keshavarzi [Muon g-2],
``{\it The Muon $g-2$ Experiment at Fermilab},''
\href{http://dx.doi.org/10.1051/epjconf/201921205003}
{EPJ Web Conf. \textbf{212} (2019), 05003}
[\href{http://arxiv.org/abs/1905.00497}{arXiv:1905.00497} [hep-ex]].

\bibitem{Fermilab21}
B. Abi et al. (Muon g - 2  Collaboration)
``{\it Measurement of the Positive Muon Anomalous Magnetic Moment to 0.46 ppm},''
\href{http://dx.doi.org/10.1103/PhysRevLett.126.141801}
{Phys. Rev. Lett. 126, 141801 (2021)}
[\href{http://arxiv.org/abs/2104.03281}{arXiv:2104.03281} [hep-ex]].

\bibitem{Bennett:2006fi}
G.~W.~Bennett \textit{et al.} [Muon g-2],
``{\it Final Report of the Muon E821 Anomalous Magnetic Moment Measurement at BNL},''
\href{http://dx.doi.org/10.1103/PhysRevD.73.072003}
{Phys. Rev. D \textbf{73} (2006), 072003}
[\href{http://arxiv.org/abs/hep-ex/0602035}{arXiv:hep-ex/0602035 } [hep-ex]].

\bibitem{Keshavarzi:2019abf}
A.~Keshavarzi, D.~Nomura and T.~Teubner,
``{\it $g-2$ of charged leptons, $\alpha (M^2_Z)$ , and the hyperfine splitting of muonium},''
\href{http://dx.doi.org/10.1103/PhysRevD.101.014029}
{Phys. Rev. D \textbf{101} (2020) no.1, 014029}
[\href{http://arxiv.org/abs/1911.00367 }{arXiv:1911.00367} [hep-ph]].

\bibitem{Hagiwara:2003da}
K.~Hagiwara, A.~D.~Martin, D.~Nomura and T.~Teubner,
``{\it Predictions for g-2 of the muon and alpha(QED) (M**2(Z))},''
\href{http://dx.doi.org/10.1103/PhysRevD.69.093003}
{Phys. Rev. D \textbf{69} (2004), 093003}
[\href{http://arxiv.org/abs/hep-ph/0312250}{arXiv:hep-ph/0312250} [hep-ph]].

\bibitem{Gerardin:2020gpp}
  A.~G\'erardin,
  ``{\it The anomalous magnetic moment of the muon: status of Lattice QCD calculations},''
  \href{http://dx.doi.org/10.1140/epja/s10050-021-00426-7}{Eur. Phys. J. A \textbf{57}, no.4, 116 (2021)}
  [\href{http://arxiv.org/abs/2012.03931}{arXiv:2012.03931} [hep-lat]].

\bibitem{Chao:2021tvp}
  E.~H.~Chao, R.~J.~Hudspith, A.~G\'erardin, J.~R.~Green, H.~B.~Meyer and K.~Ottnad,
  ``{\it Hadronic light-by-light contribution to $(g-2)_\mu$ from lattice QCD: a complete calculation},''
  \href{http://dx.doi.org/10.1140/epjc/s10052-021-09455-4}{Eur. Phys. J. C \textbf{81}, no.7, 651 (2021)}
  [\href{http://arxiv.org/abs/2104.02632}{arXiv:2104.02632} [hep-lat]].

\bibitem{Borsanyi:2020mff}
S.~Borsanyi, Z.~Fodor, J.~N.~Guenther, C.~Hoelbling, S.~D.~Katz, L.~Lellouch, T.~Lippert, K.~Miura, L.~Parato and K.~K.~Szabo, \textit{et al.}
``{\it Leading hadronic contribution to the muon 2 magnetic moment from lattice QCD},''
  \href{http://dx.doi.org/10.1038/s41586-021-03418-1}
  {Nature 593, 51–55 (2021)}
  [\href{http://arxiv.org/abs/:2002.12347}{arXiv::2002.12347} [hep-lat]].

\bibitem{Blum:2018mom}
T.~Blum \textit{et al.} [RBC and UKQCD],
``{\it Calculation of the hadronic vacuum polarization contribution to the muon anomalous magnetic moment},''
\href{http://dx.doi.org/110.1103/PhysRevLett.121.022003}
{Phys. Rev. Lett. \textbf{121} (2018) no.2, 022003}
[\href{http://arxiv.org/abs/1801.07224}{arXiv:1801.07224} [hep-lat]].

\bibitem{Blum:2019ugy}
T.~Blum, N.~Christ, M.~Hayakawa, T.~Izubuchi, L.~Jin, C.~Jung and C.~Lehner,
``{\it Hadronic Light-by-Light Scattering Contribution to the 
Muon Anomalous Magnetic Moment from Lattice QCD},''
\href{http://dx.doi.org/10.1103/PhysRevLett.124.132002}
{Phys. Rev. Lett. \textbf{124} (2020) no.13, 132002}
[\href{http://arxiv.org/abs/1911.08123}{arXiv:1911.08123} [hep-lat]].

\bibitem{Davier:2019can}
M.~Davier, A.~Hoecker, B.~Malaescu and Z.~Zhang,
``{\it A new evaluation of the hadronic vacuum polarisation contributions to the muon 
anomalous magnetic moment and to $\alpha(m_Z^2)$},''
\href{http://dx.doi.org/10.1140/epjc/s10052-020-7792-2}
{Phys. Rev. D \textbf{101} (2020) no.1, 014029}
[erratum: Eur. Phys. J. C \textbf{80} (2020) no.5, 410]
[\href{http://arxiv.org/abs/1908.00921}{arXiv:1908.00921} [hep-ph]].

\bibitem{Xiao:2017dqv}
T.~Xiao, S.~Dobbs, A.~Tomaradze, K.~K.~Seth and G.~Bonvicini,
``{\it Precision Measurement of the Hadronic Contribution to the 
Muon Anomalous Magnetic Moment},''
\href{http://dx.doi.org/10.1103/PhysRevD.97.032012}
{Phys. Rev. D \textbf{97} (2018) no.3, 032012}
[\href{http://arxiv.org/abs/1712.04530}{arXiv:1712.04530} [hep-ex]].

\bibitem{Lees:2018dnv}
J.~P.~Lees \textit{et al.} [BaBar],
``{\it Study of the reactions $e^+e^-\to\pi^+\pi^-\pi^0\pi^0\pi^0\gamma$ and 
$\pi^+\pi^-\pi^0\pi^0\eta\gamma$ at center-of-mass energies 
from threshold to 4.35 GeV using initial-state radiation},''
\href{http://dx.doi.org/10.1103/PhysRevD.98.112015}
{Phys. Rev. D \textbf{98} (2018) no.11, 112015}
[\href{http://arxiv.org/abs/1810.11962}{arXiv:1907.01556} [hep-ex]].

\bibitem{Hoferichter:2019mqg}
M.~Hoferichter, B.~L.~Hoid and B.~Kubis,
``{\it Three-pion contribution to hadronic vacuum polarization},''
\href{http://dx.doi.org/10.1007/JHEP08(2019)137}
{JHEP \textbf{08} (2019), 137}
[\href{http://arxiv.org/abs/1907.01556}{arXiv:1907.01556} [hep-ph]].

\bibitem{Benayoun:2015gxa}
M.~Benayoun, P.~David, L.~DelBuono and F.~Jegerlehner,
``{\it Muon $g-2$ estimates: can one trust effective Lagrangians and global fits?},''
\href{http://dx.doi.org/10.1140/epjc/s10052-015-3830-x}
{Eur. Phys. J. C \textbf{75} (2015) no.12, 613}.
[\href{http://arxiv.org/abs/1507.02943}{arXiv:1507.02943} [hep-ph]].

\bibitem{Jegerlehner:2009ry}
F.~Jegerlehner and A.~Nyffeler,
``{\t The Muon $g-2$},''
\href{doi:10.1016/j.physrep.2009.04.003}
{Phys. Rept. \textbf{477}, 1-110 (2009)}
[\href{http://arxiv.org/abs/0902.3360}{arXiv:0902.3360} [hep-ph]].

\bibitem{Aoyama:2020ynm}
T.~Aoyama \textit{et al.}
``The anomalous magnetic moment of the muon in the Standard Model,''
\href{doi:10.1016/j.physrep.2020.07.006}{Phys. Rept. \textbf{887}, 1-166 (2020)}
[\href{http://arxiv.org/abs/2006.04822}{arXiv:2006.04822} [hep-ph]].

\bibitem{Czarnecki:2001pv}
A.~Czarnecki and W.~J.~Marciano,
``{\it The Muon anomalous magnetic moment: A Harbinger for 'new physics'},''
\href{http://dx.doi.org/10.1103/PhysRevD.64.013014}{Phys. Rev. D \textbf{64}, 013014 (2001)}
[\href{http://arxiv.org/abs/hep-ph/0102122}{arXiv:hep-ph/0102122} [hep-ph]].

\bibitem{Yin:2020afe}
  W.~Yin and M.~Yamaguchi,
  ``{\it Muon $g-2$ at multi-TeV muon collider},''
  [\href{http://arxiv.org/abs/2012.03928}{arXiv:2012.03928} [hep-ph]];
%
\bibitem{Capdevilla:2021rwo}
  R.~Capdevilla, D.~Curtin, Y.~Kahn and G.~Krnjaic,
  ``{\it A No-Lose Theorem for Discovering the New Physics of $(g-2)_\mu$ at Muon Colliders},''
  [\href{http://arxiv.org/abs/2101.10334}{arXiv:2101.10334} [hep-ph]].

\bibitem{Lindner:2016bgg}
M.~Lindner, M.~Platscher and F.~S.~Queiroz,
``A Call for New Physics : The Muon Anomalous Magnetic Moment and Lepton Flavor Violation,''
\href{doi:10.1016/j.physrep.2017.12.001}{Phys. Rept. \textbf{731}, 1-82 (2018)}
[\href{http://arxiv.org/abs/1610.06587}{arXiv:1610.06587} [hep-ph]].

\bibitem{Gninenko:2001hx}
S.~N.~Gninenko and N.~V.~Krasnikov,
``{\it The Muon anomalous magnetic moment and a new light gauge boson},''
\href{http://dx.doi.org/10.1016/S0370-2693(01)00693-1}{Phys. Lett. B \textbf{513}, 119 (2001)}
[\href{http://arxiv.org/abs/hep-ph/0102222}{arXiv:hep-ph/0102222} [hep-ph]].

\bibitem{Baek:2001kca}
S.~Baek, N.~G.~Deshpande, X.~G.~He and P.~Ko,
``{\it Muon anomalous g-2 and gauged L(muon) - L(tau) models},''
\href{http://dx.doi.org/10.1103/PhysRevD.64.055006}{Phys. Rev. D \textbf{64}, 055006 (2001)}
[\href{http://arxiv.org/abs/hep-ph/0104141}{arXiv:hep-ph/0104141} [hep-ph]].

\bibitem{Ma:2001md}
E.~Ma, D.~P.~Roy and S.~Roy,
``{\it Gauged L(mu) - L(tau) with large muon anomalous magnetic moment and the bimaximal mixing of neutrinos},''
\href{http://dx.doi.org/10.1016/S0370-2693(01)01428-9}{Phys. Lett. B \textbf{525}, 101-106 (2002)}
[\href{http://arxiv.org/abs/hep-ph/0110146}{arXiv:hep-ph/0110146} [hep-ph]].

\bibitem{Altmannshofer:2016brv}
W.~Altmannshofer, C.~Y.~Chen, P.~S.~Bhupal Dev and A.~Soni,
``{\it Lepton flavor violating Z' explanation of the muon anomalous magnetic moment},''
\href{http://dx.doi.org/10.1016/j.physletb.2016.09.046}{Phys. Lett. B \textbf{762}, 389-398 (2016)}
[\href{http://arxiv.org/abs/1607.06832}{arXiv:1607.06832} [hep-ph]].

\bibitem{Crivellin:2010ty}
A.~Crivellin, J.~Girrbach and U.~Nierste,
``{\it Yukawa coupling and anomalous magnetic moment of the muon: an update for the LHC era},''
\href{http://dx.doi.org/10.1103/PhysRevD.83.055009}{Phys. Rev. D \textbf{83}, 055009 (2011)}
[\href{http://arxiv.org/abs/1010.4485}{arXiv:1010.4485} [hep-ph]].

\bibitem{Chen:2015vqy}
C.~Y.~Chen, H.~Davoudiasl, W.~J.~Marciano and C.~Zhang,
``{\it Implications of a light dark Higgs solution to the $g_\mu$-2 discrepancy},''
\href{http://dx.doi.org/10.1103/PhysRevD.93.035006}{Phys. Rev. D \textbf{93}, no.3, 035006 (2016)}
[\href{http://arxiv.org/abs/1511.04715}{arXiv:1511.04715} [hep-ph]].

\bibitem{Abu-Ajamieh:2018ciu}
F.~Abu-Ajamieh,
``{\it Probing Scalar and Pseudoscalar Solutions of the $g$ - 2 Anomaly},''
\href{http://dx.doi.org/10.1155/2020/175153}{Adv. High Energy Phys. \textbf{2020}, 1751534 (2020)}
[\href{http://arxiv.org/abs/1810.08891}{arXiv:1810.08891} [hep-ph]].

\bibitem{Jana:2020pxx}
  S.~Jana, V.~P.~K. and S.~Saad,
  ``{\it Resolving electron and muon $g-2$ within the 2HDM},''
  \href{http://dx.doi.org/10.1103/PhysRevD.101.115037}
  {Phys. Rev. D \textbf{101}, no.11, 115037 (2020)}
  [\href{http://arxiv.org/abs/2003.03386}{arXiv:2003.03386} [hep-ph]].

\bibitem{Marciano:2016yhf}
W.~J.~Marciano, A.~Masiero, P.~Paradisi and M.~Passera,
``{\it Contributions of axionlike particles to lepton dipole moments},''
\href{http://dx.doi.org/10.1103/PhysRevD.94.115033}{Phys. Rev. D \textbf{94}, no.11, 115033 (2016)}
[\href{http://arxiv.org/abs/1607.01022}{arXiv:1607.01022} [hep-ph]].

\bibitem{Chakraverty:2001yg}
D.~Chakraverty, D.~Choudhury and A.~Datta,
``{\it A Nonsupersymmetric resolution of the anomalous muon magnetic moment},''
\href{http://dx.doi.org/10.1016/S0370-2693(01)00419-1}{Phys. Lett. B \textbf{506}, 103-108 (2001)}
[\href{http://arxiv.org/abs/hep-ph/0102180}{arXiv:hep-ph/0102180} [hep-ph]].

\bibitem{Cheung:2001ip}
K.~m.~Cheung,
``{\it Muon anomalous magnetic moment and leptoquark solutions},''
\href{http://dx.doi.org/10.1103/PhysRevD.64.033001}{Phys. Rev. D \textbf{64}, 033001 (2001)}
[\href{http://arxiv.org/abs/hep-ph/0102238}{arXiv:hep-ph/0102238} [hep-ph]].

\bibitem{Grifols:1982vx}
J.~A.~Grifols and A.~Mendez,
``{\it Constraints on Supersymmetric Particle Masses From ($g-2$) $\mu$},''
\href{http://dx.doi.org/10.1103/PhysRevD.26.1809}{Phys. Rev. D \textbf{26}, 1809 (1982)}

\bibitem{Barbieri:1982aj}
R.~Barbieri and L.~Maiani,
``{\it The Muon Anomalous Magnetic Moment in Broken Supersymmetric Theories},''
\href{http://dx.doi.org/10.1016/0370-2693(82)90547-0}{Phys. Lett. B \textbf{117}, 203-207 (1982)}

\bibitem{Martin:2001st}
S.~P.~Martin and J.~D.~Wells,
``{\it Muon Anomalous Magnetic Dipole Moment in Supersymmetric Theories},''
\href{http://dx.doi.org/10.1103/PhysRevD.64.035003}{Phys. Rev. D \textbf{64}, 035003 (2001)}
[\href{http://arxiv.org/abs/hep-ph/0103067}{arXiv:hep-ph/0103067} [hep-ph]].

\bibitem{Stockinger:2006zn}
D.~Stockinger,
``{\it The Muon Magnetic Moment and Supersymmetry},''
\href{http://dx.doi.org/10.1088/0954-3899/34/2/R01}{J. Phys. G \textbf{34}, R45-R92 (2007)}
[\href{http://arxiv.org/abs/hep-ph/0609168}{arXiv:hep-ph/0609168} [hep-ph]].

\bibitem{Padley:2015uma}
  B.~P.~Padley, K.~Sinha and K.~Wang,
  ``{\it Natural Supersymmetry, Muon $g-2$, and the Last Crevices for the Top Squark},''
  \href{http://dx.doi.org/10.1103/PhysRevD.92.055025}
  {Phys. Rev. D \textbf{92}, no.5, 055025 (2015)}
  [\href{http://arxiv.org/abs/1505.05877}{arXiv:1505.05877} [hep-ph]].

\bibitem{Belyaev:2016oxy}
A.~S.~Belyaev, J.~E.~Camargo-Molina, S.~F.~King, D.~J.~Miller, A.~P.~Morais and P.~B.~Schaefers,
``{\it A to Z of the Muon Anomalous Magnetic Moment in the MSSM with Pati-Salam at the GUT scale},''
\href{http://dx.doi.org/10.1007/JHEP06(2016)142}{JHEP \textbf{06}, 142 (2016)}
[\href{http://arxiv.org/abs/1605.02072}{arXiv:1605.02072} [hep-ph]].

\bibitem{Endo:2019bcj}
  M.~Endo and W.~Yin,
  ``{\it Explaining electron and muon $g-2$ anomaly in SUSY without lepton-flavor mixings},''
  \href{http://dx.doi.org/10.1007/JHEP08(2019)122}
  {JHEP \textbf{08}, 122 (2019)}
  [\href{http://arxiv.org/abs/1906.08768}{arXiv:1906.08768} [hep-ph]].

\bibitem{Kpatcha:2019pve}
E.~Kpatcha, I.~Lara, D.~E.~L\'opez-Fogliani, C.~Mu\~noz and N.~Nagata,
``{\it Explaining muon $g-2$ data in the $\mu\nu$SSM},''
\href{http://dx.doi.org/10.1140/epjc/s10052-021-08938-8}{Eur. Phys. J. C \textbf{81}, no.2, 154 (2021)}
[\href{http://arxiv.org/abs/1912.04163}{arXiv:1912.04163} [hep-ph]].

\bibitem{Fayet:2007ua}
P.~Fayet,
``{\it U-boson production in e+ e- annihilations, psi and Upsilon decays, and Light Dark Matter},''
\href{http://dx.doi.org/10.1103/PhysRevD.75.115017}{Phys. Rev. D \textbf{75}, 115017 (2007)}
[\href{http://arxiv.org/abs/hep-ph/0702176}{arXiv:hep-ph/0702176} [hep-ph]].

\bibitem{Pospelov:2008zw}
M.~Pospelov,
``{\it Secluded U(1) below the weak scale},''
\href{http://dx.doi.org/10.1103/PhysRevD.80.095002}{Phys. Rev. D \textbf{80}, 095002 (2009)}
[\href{http://arxiv.org/abs/0811.1030}{arXiv:0811.1030} [hep-ph]].

\bibitem{TuckerSmith:2010ra}
D.~Tucker-Smith and I.~Yavin,
``{\it Muonic hydrogen and MeV forces},''
\href{http://dx.doi.org/10.1103/PhysRevD.83.101702}{Phys. Rev. D \textbf{83}, 101702 (2011)}
[\href{http://arxiv.org/abs/1011.4922}{arXiv:1011.4922} [hep-ph]].

\bibitem{Mohlabeng:2019vrz}
G.~Mohlabeng,
``{\it Revisiting the dark photon explanation of the muon anomalous magnetic moment},''
\href{http://dx.doi.org/10.1103/PhysRevD.99.115001}{Phys. Rev. D \textbf{99}, no.11, 115001 (2019)}
[\href{http://arxiv.org/abs/1902.05075}{arXiv:1902.05075} [hep-ph]].

\bibitem{Fabbrichesi:2020wbt}
M.~Fabbrichesi, E.~Gabrielli and G.~Lanfranchi,
``{\it The Dark Photon},''
\href{http://dx.doi.org/10.1007/978-3-030-62519-1}{The Dark Photon}
[\href{http://arxiv.org/abs/2005.01515}{arXiv:2005.01515} [hep-ph]].

\bibitem{Agrawal:2014ufa}
P.~Agrawal, Z.~Chacko and C.~B.~Verhaaren,
``{\it Leptophilic Dark Matter and the Anomalous Magnetic Moment of the Muon},''
\href{http://dx.doi.org/10.1007/JHEP08(2014)147}{JHEP \textbf{08}, 147 (2014)}
[\href{http://arxiv.org/abs/1402.7369}{arXiv:1402.7369} [hep-ph]].

\bibitem{Belanger:2015nma}
G.~B\'elanger, C.~Delaunay and S.~Westhoff,
``{\it A Dark Matter Relic From Muon Anomalies},''
\href{http://dx.doi.org/10.1103/PhysRevD.92.055021}{Phys. Rev. D \textbf{92}, 055021 (2015)}
[\href{http://arxiv.org/abs/1507.06660}{arXiv:1507.06660} [hep-ph]].

\bibitem{Kowalska:2017iqv}
K.~Kowalska and E.~M.~Sessolo,
``{\it Expectations for the muon g-2 in simplified models with dark matter},''
\href{http://dx.doi.org/10.1007/JHEP09(2017)112}{JHEP \textbf{09}, 112 (2017)}
[\href{http://arxiv.org/abs/1707.00753}{arXiv:1707.00753} [hep-ph]].

\bibitem{Calibbi:2018rzv}
L.~Calibbi, R.~Ziegler and J.~Zupan,
``{\it Minimal models for dark matter and the muon g\ensuremath{-}2 anomaly},''
\href{http://dx.doi.org/10.1007/JHEP07(2018)046}{JHEP \textbf{07}, 046 (2018)}
[\href{http://arxiv.org/abs/1804.00009}{arXiv:1804.00009} [hep-ph]].

\bibitem{Kawamura:2020qxo}
J.~Kawamura, S.~Okawa and Y.~Omura,
``{\it Current status and muon $g-2$ explanation of lepton portal dark matter},''
\href{http://dx.doi.org/10.1007/JHEP08(2020)042}{JHEP \textbf{08}, 042 (2020)}
[\href{http://arxiv.org/abs/2002.12534}{arXiv:2002.12534} [hep-ph]].

\bibitem{Jana:2020joi}
  S.~Jana, P.~K.~Vishnu, W.~Rodejohann and S.~Saad,
  ``{\it Dark matter assisted lepton anomalous magnetic moments and neutrino masses},''
  \href{http://dx.doi.org/10.1103/PhysRevD.102.075003}
  {Phys. Rev. D \textbf{102}, no.7, 075003 (2020)}
  [\href{http://arxiv.org/abs/2008.02377}{arXiv:2008.02377} [hep-ph]].

\bibitem{Kaneta:2016wvf}
K.~Kaneta, H.~S.~Lee and S.~Yun,
``Portal Connecting Dark Photons and Axions,''
\href{doi:10.1103/PhysRevLett.118.101802}{Phys. Rev. Lett. \textbf{118}, no.10, 101802 (2017)}
[\href{http://arxiv.org/abs/1611.01466}{arXiv:1611.01466} [hep-ph]].

\bibitem{Barr-Zee}
S.~M.~Barr, E.~M.~Freire and A.~Zee,
 ``{\it A Mechanism for large neutrino magnetic moments},''
\href{http://dx.doi.org/10.1103/PhysRevLett.65.2626}
{Phys. Rev. Lett. \textbf{65}, 2626-2629 (1990)};


\bibitem{Kannike:2011ng}
K.~Kannike, M.~Raidal, D.~M.~Straub and A.~Strumia,
``Anthropic solution to the magnetic muon anomaly: the charged see-saw,''
\href{doi:10.1007/JHEP02(2012)106}{JHEP \textbf{02} (2012), 106
[erratum: JHEP \textbf{10} (2012), 136]}
[\href{http://arxiv.org/abs/1111.2551}{arXiv:1111.2551} [hep-ph]].

\bibitem{Dermisek:2013gta}
R.~Dermisek and A.~Raval,
``Explanation of the Muon g-2 Anomaly with Vectorlike Leptons and its Implications for Higgs Decays,''
\href{doi:10.1103/PhysRevD.88.013017}{Phys. Rev. D \textbf{88} (2013), 013017}
[\href{http://arxiv.org/abs/1305.3522}{arXiv:1305.3522} [hep-ph]].

\bibitem{Crivellin:2018qmi}
A.~Crivellin, M.~Hoferichter and P.~Schmidt-Wellenburg,
``Combined explanations of $(g-2)_{\mu,e}$ and implications for a large muon EDM,''
\href{doi:10.1103/PhysRevD.98.113002}{Phys. Rev. D \textbf{98} (2018) no.11, 113002}
[\href{http://arxiv.org/abs/1807.11484}{arXiv:1807.11484} [hep-ph]].


\bibitem{deNiverville:2018hrc}
P.~deNiverville, H.~S.~Lee and M.~S.~Seo,
``Implications of the dark axion portal for the muon g\ensuremath{-}2 , B factories, fixed target neutrino experiments, and beam dumps,''
\href{doi:10.1103/PhysRevD.98.115011}{Phys. Rev. D \textbf{98}, no.11, 115011 (2018)}
[\href{http://arxiv.org/abs/1806.00757}{arXiv:1806.00757} [hep-ph]].

\bibitem{Alonso-Alvarez:2018irt}
G.~Alonso-\'Alvarez, M.~B.~Gavela and P.~Quilez,
``{\it Axion couplings to electroweak gauge bosons},''
\href{http://dx.doi.org/10.1140/epjc/s10052-019-6732-5}
{Eur. Phys. J. C \textbf{79} (2019) no.3, 223}
[\href{http://arxiv.org/abs/1811.05466 }{arXiv:1811.05466} [hep-ph]].

\bibitem{Bauer:2018uxu}
M.~Bauer, M.~Heiles, M.~Neubert and A.~Thamm,
``{\it Axion-Like Particles at Future Colliders},''
\href{http://dx.doi.org/10.1140/epjc/s10052-019-6587-9}
{Eur. Phys. J. C \textbf{79} (2019) no.1, 74}
[\href{http://arxiv.org/abs/1808.10323}{arXiv:1808.10323} [hep-ph]].

\bibitem{Cheung:2008ii}
K.~Cheung, T.~W.~Kephart, W.~Y.~Keung and T.~C.~Yuan,
``{\it Decay of Z Boson into Photon and Unparticle},''
\href{http://dx.doi.org/10.1016/j.physletb.2008.03.037}
{Phys. Lett. B \textbf{662} (2008), 436-440}
[\href{http://arxiv.org/abs/0801.1762}{arXiv:0801.1762} [hep-ph]].

\bibitem{Jaeckel:2015jla}
J.~Jaeckel and M.~Spannowsky,
``{\it Probing MeV to 90 GeV axion-like particles with LEP and LHC},''
\href{http://dx.doi.org/10.1016/j.physletb.2015.12.037}
{Phys. Lett. B \textbf{753} (2016), 482-487}
[\href{http://arxiv.org/abs/1509.00476}{arXiv:1509.00476} [hep-ph]].

\bibitem{Batell:2016ove}
B.~Batell, N.~Lange, D.~McKeen, M.~Pospelov and A.~Ritz,
``{\it Muon anomalous magnetic moment through the leptonic Higgs portal},''
\href{http://dx.doi.org/10.1103/PhysRevD.95.075003}{Phys. Rev. D \textbf{95}, no.7, 075003(2017)} 
[\href{http://arxiv.org/abs/1606.04943}{arXiv:1606.04943} [hep-ph]].

\bibitem{Chen:2017awl}
C.~Y.~Chen, M.~Pospelov and Y.~M.~Zhong,
``{\it Muon Beam Experiments to Probe the Dark Sector},''
\href{http://dx.doi.org/10.1103/PhysRevD.95.115005}
{Phys. Rev. D \textbf{95}, no.11, 115005 (2017)}
[\href{http://arxiv.org/abs/1701.07437 }{arXiv:1701.07437 } [hep-ph]].

\bibitem{Batell:2017kty}
B.~Batell, A.~Freitas, A.~Ismail and D.~Mckeen,
``{\it Flavor-specific scalar mediators},''
\href{http://dx.doi.org/10.1103/PhysRevD.98.055026}
{Phys. Rev. D \textbf{98} (2018) no.5, 055026}
[\href{http://arxiv.org/abs/1712.10022}{arXiv:1712.10022} [hep-ph]].

\bibitem{Croon:2020lrf}
D.~Croon, G.~Elor, R.~K.~Leane and S.~D.~McDermott,
``{\it Supernova Muons: New Constraints on $Z$' Bosons, Axions and ALPs},''
\href{http://dx.doi.org/10.1007/JHEP01(2021)107}
{JHEP \textbf{01}, 107 (2021)}
[\href{http://arxiv.org/abs/2006.13942}{arXiv:2006.13942} [hep-ph]].

\bibitem{TheBABAR:2016rlg}
J.~P.~Lees \textit{et al.} [BaBar],
``{\it Search for a muonic dark force at BABAR},''
\href{http://dx.doi.org/10.1103/PhysRevD.94.011102}
{Phys. Rev. D \textbf{94} (2016) no.1, 011102}
[\href{http://arxiv.org/abs/1606.03501}{arXiv:1606.03501} [hep-ex]].

\bibitem{Sirunyan:2018nnz}
A.~M.~Sirunyan \textit{et al.} [CMS],
``{\it Search for an $L_{\mu}-L_{\tau}$ gauge boson using Z$\to4\mu$ events in proton-proton collisions at $\sqrt{s} =$ 13 TeV},''
\href{http://dx.doi.org/10.1016/j.physletb.2019.01.072}
{Phys. Lett. B \textbf{792} (2019), 345-368}
[\href{http://arxiv.org/abs/1808.03684}{arXiv:1808.03684} [hep-ex]].

\bibitem{Filippi:2020kii}
  A.~Filippi and M.~De Napoli,
  ``{\it Searching in the dark: the hunt for the dark photon},''
  \href{http://dx.doi.org/10.1016/j.revip.2020.100042}
  {Rev. Phys. \textbf{5}, 100042 (2020)}
  [\href{http://arxiv.org/abs/2006.04640}{arXiv:2006.04640} [hep-ph]].

\bibitem{Lees:2014xha}
J.~P.~Lees \textit{et al.} [BaBar],
``{\it Search for a Dark Photon in $e^+e^-$ Collisions at BaBar},''
\href{http://dx.doi.org/10.1103/PhysRevLett.113.201801}
{Phys. Rev. Lett. \textbf{113} (2014) no.20, 201801}
[\href{http://arxiv.org/abs/1406.2980}{arXiv:1406.2980} [hep-xp]].

\bibitem{Batley:2015lha}
  J.~R.~Batley \textit{et al.} [NA48/2],
  ``{\it Search for the dark photon in $\pi^0$ decays},''
  \href{http://dx.doi.org/10.1016/j.physletb.2015.04.068}
  {Phys. Lett. B \textbf{746}, 178-185 (2015)}
  [\href{http://arxiv.org/abs/1504.00607}{arXiv:1504.00607} [hep-ex]].

\bibitem{Ilten:2016tkc}
  P.~Ilten, Y.~Soreq, J.~Thaler, M.~Williams and W.~Xue,
  ``{\it Proposed Inclusive Dark Photon Search at LHCb},''
  \href{http://dx.doi.org/10.1103/PhysRevLett.116.251803}
  {Phys. Rev. Lett. \textbf{116}, no.25, 251803 (2016)}
  [\href{http://arxiv.org/abs/1603.08926}{arXiv:1603.08926} [hep-ph]].

\bibitem{Aaij:2019bvg}
R.~Aaij \textit{et al.} [LHCb],
``{\it Search for $A'\to\mu^+\mu^-$ Decays},''
\href{http://dx.doi.org/10.1103/PhysRevLett.124.041801}
{Phys. Rev. Lett. \textbf{124} (2020) no.4, 041801}
[\href{http://arxiv.org/abs/1910.06926 }{arXiv:1910.06926 } [hep-ex]].

\bibitem{Sirunyan:2019wqq}
  A.~M.~Sirunyan \textit{et al.} [CMS],
  ``{\it Search for a Narrow Resonance Lighter than 200 GeV Decaying to a Pair of Muons in Proton-Proton Collisions at $\sqrt{s} =$  TeV},''
  \href{http://dx.doi.org/10.1103/PhysRevLett.124.131802}
  {Phys. Rev. Lett. \textbf{124}, no.13, 131802 (2020)}
  [\href{http://arxiv.org/abs/1912.04776}{arXiv:1912.04776} [hep-ex]].

\bibitem{Curtin:2014cca}
D.~Curtin, R.~Essig, S.~Gori and J.~Shelton,
``{\it Illuminating Dark Photons with High-Energy Colliders},''
\href{http://dx.doi.org/10.1007/JHEP02(2015)157}{JHEP \textbf{02} (2015), 157} 
[\href{http://arxiv.org/abs/1412.0018}{arXiv:1412.0018} [hep-ph]].

\end{thebibliography}
\end{document}